\documentclass[twocolumn,showpacs,amsmath,amssymb,prd,floatfix,preprintnumbers]{revtex4}
\usepackage[dvipdfmx]{graphicx}
\usepackage{bm}
\usepackage{amsmath}
\usepackage{color}
\usepackage{xcolor}
\usepackage{hyperref}
\usepackage{comment}
\usepackage{multirow}    
\usepackage[varg]{txfonts}

\hypersetup{
  bookmarksnumbered=true,
  linkcolor=red,
  citecolor=green,
  urlcolor=cyan
}

\newcommand{\simgt}{\lower.5ex\hbox{$\; \buildrel > \over \sim \;$}}
\newcommand{\simlt}{\lower.5ex\hbox{$\; \buildrel < \over \sim \;$}}

\begin{document}

\title[]{Redshift drift  in radially inhomogeneous Lema\^itre-Tolman-Bondi spacetimes}

\author{Romain Codur }
\author{Christian Marinoni }
\email{christian.marinoni@cpt.univ-mrs.fr}
\affiliation{Aix Marseille Univ, Universit\'e de Toulon, CNRS, CPT, Marseille, France}

\date{\today}

\begin{abstract}
We provide a formula  for estimating the redshift and its secular change (redshift drift) in  Lema\^itre-Tolman-Bondi (LTB) spherically symmetric universes. We compute the scaling of the redshift drift for LTB models that predict Hubble diagrams indistinguishable from those of the standard cosmological model, the flat lambda cold dark matter ($\Lambda$CDM) model. We show that the redshift drift for these degenerate LTB models is typically different from that predicted in the $\Lambda$CDM scenario.
We also highlight and discuss  some unconventional redshift-drift signals that arise in LTB universes and give them distinctive features compared to the standard model.
We argue that the redshift drift is a metric observable  that allows us to reduce the degrees of freedom of spherically symmetric models and to make them more predictive and thus falsifiable.

\end{abstract}


\maketitle

\section{Introduction}
\label{intro}

The cosmological principle (CP) states that the universe is homogeneous and isotropic on large scales, i.e., invariant under translations and rotations around each comoving observer.
Long regarded  only as a philosophically appealing, if not logically necessary, convention on the symmetries of space
\cite{Milne1935,Harrison2000}, it has now acquired the most scientifically legitimate status of a useful working hypothesis that can be questioned and verified by astronomical data \cite{Maartens2011, Uzan2008, Marinoni2012, Valkenburg2014, Zhang2015, Park2017, Sarkar2019, Jimenez_2019, Camarena2021}. 
The fact that the standard  cosmic metric deduced from it, the Friedmann-Robertson-Walker line element (FRW),  is still practically indisputable, should not make us lose sight of the importance of challenging the standard paradigm and analysing the cosmological relevance of alternative spacetime proposals.  

The  phenomenology induced by cosmic  inhomogeneities and their impact on the interpretation of past light cone data  is still far from being satisfactorily understood   \cite{Krasinski1997}.
Nonetheless, among the various  inhomogeneous spacetimes,   spherically symmetric and radially inhomogeneous  models of the universe, the so-called Lema\^itre-Tolman-Bondi (LTB) models \cite{Lemaitre1933,Tolman1934,Bondi1947} have being investigated in some details.  They are widely regarded  as interesting testing ground for challenging the CP  or for non-canonical  
interpretations of  cosmological observations  \citep[see][for a review]{Enqvist2008, Marra2011}.   

The popularity of the LTB metric models stems from the fact that they  offer an alternative physical explanation for the Hubble diagram measured via Supernovae Ia (SNIa).
The  weakening of light from distant supernovae would be caused, in this inhomogeneous scenario, by radial spatial gradients in the rate of expansion and density of matter, rather than, 
as currently preferred, by a dark energy component forcing the second derivative of the cosmic scale factor $a(t)$ to be positive \cite{Celerier1999, Tomita2000,  Enqvist2007, Bellido2008a, Biswas2010, Redlich2014, Stahl2016, Tokutake2018}. Indeed, since  we have direct access only to data on our light-cone,  it is challenging  to disentangle temporal evolution in
the scale factor from radial spatial variations. For example,  it is possible to mimic the effect of dark energy 
by postulating that we live in a region of the universe underdense with respect to the average,  which  extends to the point in space where the acceleration/deceleration transition takes place (at redshift $z\sim 0.6$).  Typically one invokes the hypothetical existence of a deep void, almost centered around us, with a radius that, depending on models,  can be as big as the horizon ($\sim 3$ Gpc). Interestingly,   LTB  modelling of smaller size voids ($\sim 300$Mpc), which are  compatible with tentative evidences about the existence of a local under density in the galaxy distribution
\cite{Keenan2013, Shanks2019}, has been recently advocated  as a  way to lessen the tension between local and early epochs measurements of the
present day Hubble parameter $H_0$ \cite{Tokutake2018,Lukovic2020}.

From the theoretical side, it has been  argued that  any spatially uniform data set can be misinterpreted by carefully tuning the functional degrees of freedom of an inhomogeneous  LTB model \cite{Mustapha1997}. This is ultimately due to the fact that   this spacetime has fewer symmetries than the standard metric of the universe. For example, 
distance-redshift relations of the standard $\Lambda$CDM  models can  be reconstructed  with arbitrary precision using the freedom in the specification of the arbitrary boundary conditions of the LTB models \cite{Sundell2015}.
Even if fine-tuned to reproduce the $\Lambda$CDM Hubble diagram,  the LTB models  should retain their cosmological predictiveness when distance-independent observables are considered.  It is thus of some interest to devise cosmological testing schemes that might resolve this degeneracy.  In this paper we explore if and to what level this can be achieved by studying the temporal drift of the  
redshift \cite{Sandage1962,Loeb1998}.

The time variation of the redshift  is a cosmological quantity  that, being directly measurable from data, gives access to  the evolution of the  past-light cone in real time  and in a  model-independent way \cite{Kim2015}. It is thus a crucial  observable  for  the  so-called  cosmographic  approach  to cosmology \cite{Visser2004}.
While the physics is straightforward, the observational implementation of this probe  is challenging: dimensional  arguments suggests that the signal is of order $\dot{z}\sim H(t)$ and thus 
very precise and stable observations over a period of several years  are necessary to  reach a signal to noise ratio of unity.
Notwithstanding,  the Extremely Large Telescope (ELT)  \cite{Liske2008,Martinelli2012}, the Square Kilometre Array (SKA) \cite{Klockner2015} and the Canadian Hydrogen Intensity Mapping Experiment (CHIME) 
\cite{Yu2014}  observational facilities promise to  reach the necessary sensitivity.

Redshift drift data, once interpreted  in the  FRW framework,   are expected  to 
provide competitive constraints on  the  Hubble constant \cite{Corasaniti2007} as well as on  some representative dark energy models  \cite{Balbi2007, Zhang2007, Geng2014, Denkie2014, Martins2016, Guo2016,Lazkoz2018, Alves2019}.
The scaling of the redshift drift has also been investigated in alternative metric scenarios such as in arbitrary spacetimes \cite{Korzy_ski_2018, Heinesen_2021}, 
in  Stephani universes \cite{Balcerzak2013},  in axially symmetric quasi-spherical Szekeres models \cite{Mishra2012}
and backreaction models \cite{Koksbang2016}. 
As far as the LTB model is concerned, the redshift drift  has been calculated numerically by \cite{Yoo2011} although,  as argued in \cite{Quercellini2012},   the intrinsic smallness of the redshift drift effect is a challenging factor for numerical codes, demanding  careful architecture of non-na\"ive algorithms.
An exact formula  using observational coordinates has been proposed by \cite{Uzan2008} although it does not match numerical results, as claimed by \cite{Koksbang2016}. 
Here we explore these issues and present an exact algebraic expression that provides the explicit dependence of the redshift drift on the structural parameters of the LTB metric
in comoving coordinates.

The paper is organized as follows;
in Sec.~\ref{sec:LTBmodel}, we briefly review some key features of the LTB cosmological model and introduce our notations. 
In Sec.~\ref{sec:zco}, we derive an explicit expression for the redshift  of light as a function of the LTB comoving coordinates, as seen by a an observer 
sitting at the center of symmetry of the metric while in \ref{sec:zdco} we  provide the analytical expressions of its 
first and second time variations. 
Various potential systematic biases  are discussed in  Sec. \ref{sec:bias} and applications to a specific  LTB model are  shown in Sec. \ref{sec:LTBexplicit}.
Sec.~\ref{sec:conclusion} provides summary and conclusion.

In the following,  we present results  in natural units  ($c=1$) and we refer 
to the standard  $\Lambda$CDM
model, as the  flat FRW  spacetime which best fits the Planck18 data \cite{Planck2018}.

\section{ The LTB cosmological model}  \label{sec:LTBmodel}
\subsection{Kinematics of light}
Universes with  radial  spatial inhomogeneities can be charted using 
spherical  coordinates $x^i \equiv (\chi, \theta, \phi) $  that   comove ( $U^i\equiv dx^i / dt = 0$ ) with matter. 
It is convenient to  set up the spatial origin ($x^i = 0$), the center of symmetry,  at the observer position, and to 
choose the  time coordinate ($x^0 \equiv  t$) so that it  measures the proper time of the comoving matter particles.
The  line element   takes the form 
\begin{equation}
 ds^2=dt^2-\alpha^2(t, \chi) d\chi^2 -A^2(t, \chi)(d\theta^2+\sin^2 \theta d\phi^2)
\label{dsltb}
 \end{equation}
and we will refer to the two degrees of freedom  $\alpha$ and $A$ as the radial  and angular scale factor, respectively. 

The geodesic motion of massive and massless particles  for the central  LTB observer is given by 
\begin{align}
\frac{d k^{\mu}}{d\lambda}+\Gamma^{\mu}_{\nu \rho}k^{\nu} k^{\rho}=0
\end{align}
where $\mu=0,1,2,3$, $k^{\mu}=dx^{\mu}/d\lambda$  is the tangent to the null geodesics  and $\lambda \in \mathbb{R}$ is an affine parameter that describes the particles path.
Because of the spherical symmetry, the geodesics are entirely determined by only two equations 
 \begin{align}
 \frac{dk^0}{d\lambda}& =-\alpha \dot{\alpha}(k^1)^2
     \label{offko} \\
 \frac{dk^1}{d\lambda}& =-\frac{\alpha'}{\alpha}(k^{1})^2-2\frac{\dot{\alpha}}{\alpha}k^0k^1  \label{offk1}
 \end{align}
where the overdot and a prime indicate  a partial derivative with respect to the coordinates  $t$ and $\chi$ respectively. 
Null-geodesics are further characterized by   the constraint $k^{\mu}k_{\mu}=0$, which supplies the additional condition
\begin{align}
 k^0=-\alpha k^1
\end{align}
where the sign convention stems from the hypothesis that the affine parameter and the radial coordinate $\chi$ are chosen to increases in the opposite direction of cosmic time, 
\textit{i.e.} the equations are  integrated backward on the past light cone of the observer. Moreover, since the affine  parameter is determined up to 
a linear transformation, without loss of generality we set $\lambda=0$ at the observer position and set as boundary condition $k^0(\lambda=0)=-1$ in the chosen 
length units.

\subsection{Dynamics of dust}\label{sec:dynamics}
The evolution of $\alpha$ and $A$ follows from 
imposing that the metric  evolves in space and  time according to the Einstein's field equations (EFE). For the purpose of our analysis, which deals with 
 low-redshift phenomena,  we will assume that the  
universe is in a dust dominated phase. Therefore the stress energy tensor  describing the matter distribution, modelled as a perfect fluid,  is  $T^{\mu \nu } = \rho_m(t, \chi)U^{\mu} U^{\nu}$ 
where $U^{\mu}=\delta^{\mu}_0$ are the components of the 4-velocity-field of the fluid with respect to the chosen coordinate system and $\rho$ 
the invariant matter density seen by a comoving observer.

The EFE, supplemented by the cosmological member $\Lambda$, remove one metric degrees of freedom by imposing 
 the constraint  
 \begin{align}
 \alpha = A'  / \sqrt{1-k(\chi)}
 \end{align}
where $k(\chi) <1$ is an arbitrary function 
 of the radial coordinate $\chi$  only,  and a prime denotes a derivative with respect to this coordinate. This function 
 describes the position dependent spatial curvature of the $t=const$ hypersurfaces.
 The 
function $A$ is determined via the equation
\begin{equation}
\left( \frac{\dot{A}} {A }\right)^2+\frac{k}{A^2}=\frac{8 \pi G}{3}  \left( \tilde{\rho}+\rho_{\Lambda}\right)
\label{ltb1}
\end{equation}
where $\rho_{\Lambda}$ is the non-dilutive energy density 
associated with the cosmological constant $\Lambda$ and where we have defined the {\it flat average density} of matter 
\begin{equation} 
\tilde{\rho}(t,  \chi) =3\; \frac{ \int_{0}^{\chi} \rho_m  A^2 A' d\chi} {A^3}
\label{md}
\end{equation}
as the  non-local  quantity satisfying the continuity equation (see Appendix A for details)
\begin{equation}
\dot{\tilde{\rho}}+3\frac{\dot{A}}{A}\tilde{\rho}=0.
\label{icont}
\end{equation}
This quantity must not be conceptually confused with (and it is in general quantitatively different from) the actual average density inside a shell of comoving coordinate $\chi$. 
Indeed,  the averaging procedure  is not carried out over the  (spherical) volume element $dV=4\pi A^2 A' (1-k(\chi))^{-1/2} d\chi $. 
Anyway, as the name suggests,  the flat and the actual average  densities of matter coincide  in LTB models with $k(\chi)=0.$
Moreover, in the absence of inhomogeneities,  when the average density coincides with the local one, one finds that  
$\tilde{\rho}(t) =\rho_m(t)$.  By inverting Eq.  (\ref{md})  one also finds the expression of the local matter density as a function of $\tilde{\rho}$
\begin{equation}\label{eq:rhoti}
    \rho_m = \Tilde{\rho} + \Tilde{\rho}' \frac{A}{3A'}.
\end{equation}
The advantage of introducing $\tilde{\rho}$  is that it 
brings the LTB formulas into closer formal analogy with those of the standard model of cosmology. This is shown explicitly by the equations (\ref{ltb1}), (\ref{icont}) or 
(\ref{omegam}) and is also evident by inspecting the acceleration equation, which, in this notation, reads simply 
\[\frac{\ddot{A}}{A}=-\frac{4}{3} \pi G \left( \tilde{\rho} - 2 \rho_{\Lambda} \right).\]
Finally, note that if there is no shell crossing during the motion of the spherically symmetric matter inhomogeneity, the integral on the numerator of the 
{\it rhs} of Eq. (\ref{md}) is time independent.

Since Eq. \ref{ltb1} is  valid at any time, by evaluating it, for example,  at present time $t_0$ we can determine $k(\chi)$ as 
\[ k(\chi)  = H_0(\chi) ^2 A_0(\chi)^2 \left( \Omega_{m 0} (\chi)+\Omega_{\Lambda 0}(\chi) -1 \right)\]
where $H_0(\chi) \equiv H(t_0, \chi) $ and $\Omega_{m 0}(\chi) \equiv \Omega_m (t_0, \chi)$  are the present day value of the (transverse)  
expansion rate 
\[H(t, \chi) \equiv \frac{\dot{A}}{A} \]
and of the density parameter 
\begin{equation}
\Omega_m(t, \chi) \equiv \frac{8\pi G}{3 H^2} \tilde{\rho}
\label{omegam}
\end{equation}
respectively.
The curvature function  $k(\chi)$ rests,  in a degenerate way,  on two unknown gauge functions $H_0(\chi)$ and $A_0\equiv A(t_0,\chi)$. We can  conveniently remove this degeneracy   by arbitrarily  fixing  the angular scale $A_0(\chi)$.  This must be a smooth and invertible positive function, and in what follows we choose the conventional gauge of the literature and set 
$A_0(\chi)=\chi$. In other terms, we label each dust shell by means of its present day areal radius. 
Once a given matter density $\rho_m$ (or equivalently $\Omega_{m 0}(\chi)$) and a present day transverse  expansion rate $H_0(\chi)$ are chosen,   
Eq. \eqref{ltb1} can be solved to determine the remaining degree of freedom $A(t, \chi).$

\section{Redshift} \label{sec:zco}
The redshift is defined  in terms of the photon wave-vector as \cite{Kristian1966}
\begin{equation}
1+z=\frac{(u_{\alpha}k^{\alpha})_{e} }{(u_{\alpha}k^{\alpha})_{0}}
\label{zdefe}
\end{equation}
where the suffix $e$ and $0$ indicate quantities that are computed at emission and present time respectively.
The wave-vector components satisfy $k^{\alpha} k_{\alpha}=0$ and  $k^0$, the physical energy of the photon,  depends at most on $t$ and $\chi$ because of the rotational symmetries of the line element. 
We thus deduce that 
\begin{equation} 
 1+z=\frac{k^0(t,\chi)}{k^0(t_0,0)}.
 \end{equation}
 where we have set $\chi_e=\chi$,  $\chi_0=0$ and $t_e=t$.
The spatial and temporal  evolution of the photon energy is accounted for  by the  geodesic equation for massless particles 
(\ref{offko}),   which can be recast  as the following partial differential equation
\begin{equation}
\frac{\partial (\alpha  k^0)}{\partial t}  = \frac{\partial k^0}{\partial \chi}.
\label{pde}
\end{equation}
By changing variable $U(t, \chi)=\alpha(t, \chi) k^0(t, \chi)$  and supposing that it exists a characteristic curve  
$t=t(\chi)$ on which  the function $U(t(\chi),\chi)$ satisfy an ordinary differential equation, Eq. (\ref{pde})  becomes 
\begin{equation}
 \frac{d U}{d \chi}-\frac{\alpha'}{\alpha}U=
 \frac{\partial U}{\partial t} \left( \alpha +  \frac{dt}{d\chi}  \right)
  \label{a1}
 \end{equation}
The equation is satisfied if 
\begin{equation}
\frac{dt}{d\chi}=-\alpha(t, \chi),
\label{geode}
\end{equation}
which shows that the characteristic curve is indeed the null  geodesic of the LTB metric,  and 
\begin{equation}
\frac{dU}{d\chi}=\frac{\alpha'}{\alpha}U.
\end{equation}
It follows that  the general  solution, which depends on an arbitrary function $g,$  fixed by the initial data, is 
\[U(t, \chi)=g \left (\chi+\int\frac{dt}{\alpha} \right )e^{\int \frac{\alpha'}{\alpha}d\chi} \]
and therefore the 
redshift is 
\begin{equation}
 1+z=\frac{\alpha(t_0, 0)}{\alpha( t(\chi), \chi)} e^{ \textstyle  \int_{0 }^{\chi}  \frac{\alpha'}{\alpha}d\chi}, 
 \label{zf}
 \end{equation}
where the integral is performed along the photon path $t=t(\chi,t_0)$ solution of (\ref{geode}). 

The previous formula makes explicit  the fact that the redshift 
depends not only on the time of emission and reception of the photon  but also on the specific geodesic path of the photon $t=t(\chi)$.
In the limiting case of a FRW model ($\alpha=a$ and $\alpha'=0$) the redshift  expression converges to the standard model one $1+z=a(t)/a(t_0)$,
with the change in frequency  resulting from the relative motion of  freely falling matter. 
In the LTB model,  instead, the varying gravitational potential between the observer and the emitter also 
contributes to the relative change in the photon frequency. As a result  the redshift depends on the integrated variation of the gravitational field along the photon null geodesics.

Expression (\ref{zf} ) shows that, under particular circumstances,  the  redshift could be a non-monotonic function of the look-back time or  of the radial coordinate $\chi$.
Photons emitted at two different radial positions, or two different times, might be characterized by  the same  redshift.
Indeed, the argument of the integral is not necessarily positive, and indeed it may change sign when the radial scale factor $\alpha$ is a non monotonic 
function of $\chi$ (while still being monotonic in the $t$ variable). 
 This might happen  even if  the scale factor $A$ is a monotonic function of  both temporal and  radial coordinates.
A distinctive  consequence of this phenomenology will be presented in Sec. \ref{sec:LTBexplicit}.

For small physical separations $\delta \chi$ between the light source and the receiver,  a series expansion  to leading order of (\ref{zf})
results in a formula  similar to the Hubble-Lema\^itre expansion law, {\it i.e.} 
\begin{equation}
 z \approx  \frac{\dot{\alpha}_0}{\alpha_0 } r 
 \end{equation}
where $ r \equiv \alpha_0 \delta \chi.$  
It is straightforward to show that $\dot{\alpha}_0/\alpha_0=\dot{A}_0/A_0$, however for $ t\ne t_0$ or $\chi \ne 0$
 the ratio $\dot{\alpha}/\alpha$
does not coincide with   $H=\dot{A}/A$, which  encapsulates information on how different shells of matter moves under the action of gravity.  To emphasise this we define the longitudinal Hubble function as
\[H_{\parallel} \equiv \frac{\dot{\alpha}}{\alpha}.\]

Finally, by taking total derivatives of  (\ref{zf})   with respect the $t$ and $\chi$, one  recovers standard expressions  quantifying how the  redshift change  when either the  photon-flight time or the 
comoving coordinate interval between source and receiver vary. The equations that specify the past null cone are
\begin{align}
\frac{dz}{dt} & =-(1+z) H_{\parallel} \label{dzdt} \\
\frac{dz}{d\chi} & =  (1+z) \alpha H_{\parallel}  \label{dzdchi} 
\end{align}
respectively. These are the pair of coupled differential equations that 
determine the relations between the coordinates and the redshift, {\it i.e.} $t(z)$ and $r(z)$.

\section{Redshift drift}\label{sec:zdco}
Now consider a pair of photons that are both emitted and received by the same comoving objects at radial coordinates  $\chi$ and $\chi_0=0$, respectively. 
Suppose that they are  emitted(/received) at two instants $t$ and $t+\delta t$(/$t_0$ and $t_0+\delta t_0$), which are separated by a time interval over which  
cosmological changes cannot be anymore overlooked (typically on timescales of the order of a few years).

Formula \ref{zf} tells us that the  redshift of a photon  is formally a functional  $z=z(t_0, t, \chi(t)) $ which depends on the specific geodesic path 
of the photon $\chi=\chi(t)$, on top of the integration boundaries $t$ and $t_0$. At the same coordinate time $t'$, two different photons, emitted at two different times by the same comoving source,  will be located at different radial coordinates  $\chi'$ and $\chi'+\delta \chi$. If these two photons, emitted with a time delay $\delta t$ are 
detected by the comoving observer with a time lag $\delta t_0$, then the difference in their  redshifts  can be determined by computing the following  variational quantity
\[
\delta z=z(t_0+\delta t_0,  t+\delta t, \chi(t)+\delta \chi(t))-z(t_0,  t,  \chi(t)).
\label{defzz}
\] 
or, in first order approximation, as 
\begin{equation}
\delta z=\frac{\partial z}{\partial t} \delta t+\frac{\partial z}{\partial t_0}\delta t_0 + (1+z)  \int_t^{t_0}dt' \delta \chi(t')  \frac{\delta z }{\delta \chi(t')}
\label{dz}
\end{equation}
where   $\delta z/\delta \chi(t')$ is the functional derivative of the redshift which, using \ref{zf}, evaluates to 
\begin{equation}
\frac{\delta z}{\delta \chi(t')}=(1+z)\partial_{\chi}H_{\parallel}(t', \chi(t')). 
\end{equation}
Since, by  definition,  time dilations are related as (see Appendix B for a detailed discussion)
\begin{equation}
\frac{\delta t_0}{\delta t}=(1+z)
\label{dilations}
\end{equation}
by plugging the expressions \ref{zf}  and \ref{dilations} into Eq. \ref{dz} we obtain 
\begin{eqnarray}
\frac{\delta z}{\delta t_0} &  = &  (1+z) \int_t^{t_0} \frac{\partial_{\chi} H_{\parallel}(t',\chi(t')) dt'}{\alpha(t',\chi(t'))(1+z(t',\chi(t')))} \nonumber  \\
& & +H_{\parallel}(0,t_0) (1+z)-H_{\parallel}(\chi(z),t(z))
\label{zd}
\end{eqnarray}
where $t(z)$ and $\chi(z)$ are calculated along the ray path of the photon. As for the redshift [see Eq. \ref{zf}], the  redshift drift is also  independent from the transverse scale factor  $A$ and uniquely depends on the longitudinal Hubble function.
Moreover, note that Eq. (\ref{zd})   reduces to the standard FRW formula when radial inhomogeneities vanish.

Up to this point, the discussion has relied solely on kinematic concepts, with no reference at all to the dynamical evolution of the metric. Indeed, Eq. 
(\ref{zd}) can be used to predict redshift drift in any spherically symmetrical spacetime regardless of whether its evolution  satisfies Einstein's equations.
However, if the metric changes in time according to the standard field equations of general relativity   (see Sec. \ref{sec:dynamics}),  the degrees of freedom of the LTB metric
 can be reduced. The price to pay is that   the redshift drift becomes {\it model dependent}. The gain is that its redshift dependence can be easily 
 predicted since the redshift drift becomes a function of the transverse expansion rate $H$. 
The model-dependent formula for the redshift drift can be obtained by simply replacing $\alpha=A'(\chi)$  and 
\[ 
H_{\parallel}=H   \left ( 1+   \frac{(\log H)'}{(\log A)'} \right ). 
\]
in (\ref{zd}).

The cosmological relevance  of the rate of change of the redshift drift itself, i.e. the next order time derivative of the redshift drift,   has been pointed out by 
\cite{Lake2007}, and   \cite{Martins2016} have provided illustrations of the power of this observable  in discriminating different cosmological models. 
These authors also  emphasizes  that such a measurement is well within the reach of the
SKA Phase 2 array as well as of ELT-HIRES (albeit, in this case, with less sensitivity). A complementary goal 
of this article is therefore also to predict the amplitude and scaling of  the second time derivative of the redshift within the framework of LTB cosmologies. 

By proceeding as before, and taking the variation  of the redshift drift formula,  we can predict the amplitude and scaling of  the second time derivative of the redshift within the framework of LTB cosmologies. The rate at which the drift changes is given by 

\begin{eqnarray}
\frac{\delta^2 z}{\delta t_0^2} & = & (1+z)\left[\frac{dH_{\parallel}}{dt_0}+g_0+\int_{t}^{t_0} \frac{\partial_{\chi} g}{\alpha (1+z)} dt' \right] + \\
 & & (1+z)^{-1}\left[ \left(\frac{\delta z}{\delta t_0}\right)^2-H_{\parallel}\left( H_{\parallel 0}+\frac{\delta z}{\delta t_0}\right)+\frac{dH_{\parallel}}{dt} \right]-H_{\parallel 0}H_{\parallel}
\nonumber 
\end{eqnarray}
where 
\[ g(t_0,t,  \chi(t))= \frac{\partial_{\chi} H_{\parallel}(t,\chi(t)) dt}{\alpha(t,\chi(t))(1+z(t_0,t, \chi(t)))}\]
\[\frac{dH_{\parallel}}{dt}=\frac{\partial H_{\parallel}}{\partial t}+\frac{\partial H_{\parallel}}{\partial \chi} \frac{d\chi}{dt_0}= \dot{H}_{\parallel}-\frac{H'_{\parallel}}{\alpha} \]
and where a suffix $0$ means that the quantities are evaluated at $\chi=0$ at present time $t_0$.

Again,  in the limit in which the relevant quantities do not depend on the radial coordinate $\chi,$   the standard FRW formula 
\[\frac{\delta^2 z}{\delta t_0^2} = \dot{H}_0(1+z)+H_0^2(1+z)-H_0H-\frac{\dot{H}}{1+z}\]
is recovered.

\section{Systematic effects}\label{sec:bias}
Here we investigate potential systematics that might perturb the redshift drift measurement. 
On top of numerical artifacts that arise because of the different approximation schemes with which the  redshift drift is defined and computed, 
we also analyse the impact of physical effects, such as proper motions of the sources or the observer.

\subsection{Redshift drift definition}  

There are three possible way in which the redshift drift can be defined.
The definition that most closely capture the essence of the phenomenon is  \ref{defzz} 
where the redshift is computed using the formula (\ref{zf}) [and the time dilation using Eq.  (\ref{dilations})]. No approximations are involved in this definition that we refer to as 
 {\it method} 1.

A second way of defining the redshift drift is via the linear approximation Eq. (\ref{dz}) applied to the integral solution \ref{zf}.
This approach, here called {\it method 2}, captures leading order contributions to the time change of the redshift,  and leads to the explicit formula \ref{zd}.
Conceptually,  this approach differs
from the approximation scheme, applied almost ubiquitously in literature, which consists in linearizing the 
geodesic equations [cf. eqs. (\ref{dzdt}) and (\ref{dzdchi})]  and computing, numerically, the redshift drift. According to this last strategy ({\it method 3}), 
 the amplitude of the drift follows   from solving the approximated  differential equation 
\begin{equation}
\frac{d}{d \lambda} \delta z (\lambda)=\frac{\partial \chi}{\partial \lambda} \dot{\alpha} \delta z+(1+z) \ddot{\alpha} \delta z(\lambda)
\end{equation}
with the initial condition $\delta z(0)=0$ along the geodesic path of the photon $\chi(\lambda), t(\lambda)$ (see Appendix C). 

\begin{figure}
\begin{center}
	\includegraphics[scale=0.41]{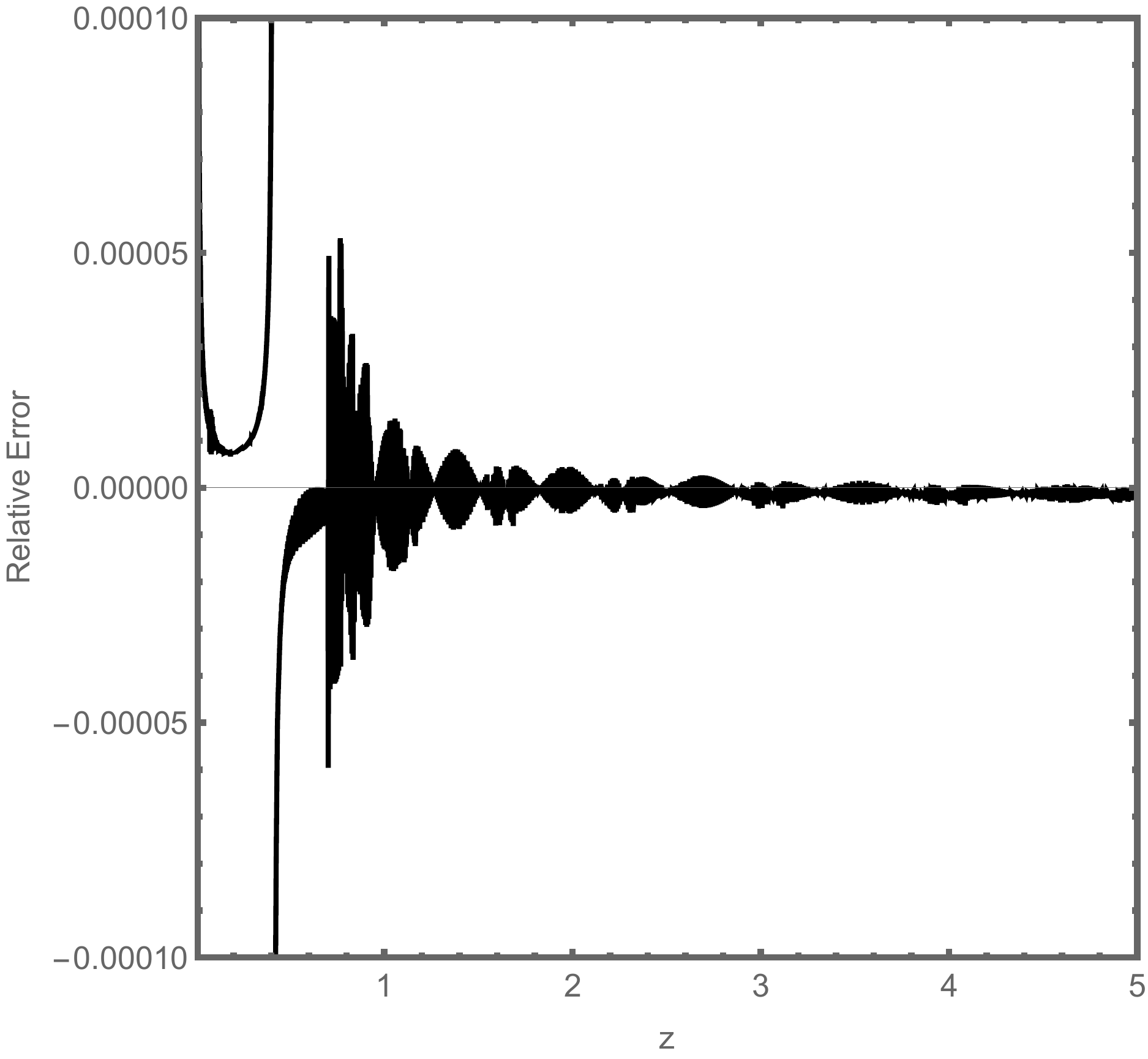}
			\caption{ Relative error in the  redshift drift estimated according to models 1 and 2. The relative error is calculated as (model 1 - model 2)/(model 2).  We have assumed  the $\Lambda$LTB universe  as reference cosmology (see Sec. \ref{sec:LTBexplicit}).}
				\label{Fig_num}
		\centering
	\end{center}
\end{figure}

Given the explicit  solution for the redshift in LTB cosmologies given in Eq. (\ref{zf}), we are in measure to compare all these definitions, check their 
mutual consistency and thus the effectiveness of the approximations involved. The results of applying these schemes to the $\Lambda$LTB model
is shown in Fig. \ref{Fig_num}.

Errors due to the linear approximation involved in both methods 2 and 3 as well as the numerical imprecision in computing the geodesics path of the photons (via equations 
(\ref{dzdt}) and (\ref{dzdchi})) are negligible, with the relative imprecision of both methods 2 and 3 with respect to method 1 of order 
$\sim 5 \times 10^{-5}$ over all the redshift range explored.  The ``theoretical" noise  is  statistically insignificant, if compared to the ``observational" imprecisions expected for the measurements of the redshift drift signal, and there is no need  to devise a more refined analytical way for computing the redshift drift nor a more stable  numerical way to solve for the geodesics in LTB cosmologies.

The relative imprecision  in computing the redshift drift according to methods 2 and 3
is even smaller, of order $\sim 10^{-7}$. This insignificant  residual
discrepancy  results from the  linearisation procedure being applied to the  differential equations before integration (method 3)  or directly  to the integral expression for the redshift (method 2).

\begin{figure}
\begin{center}
	\includegraphics[scale=0.76]{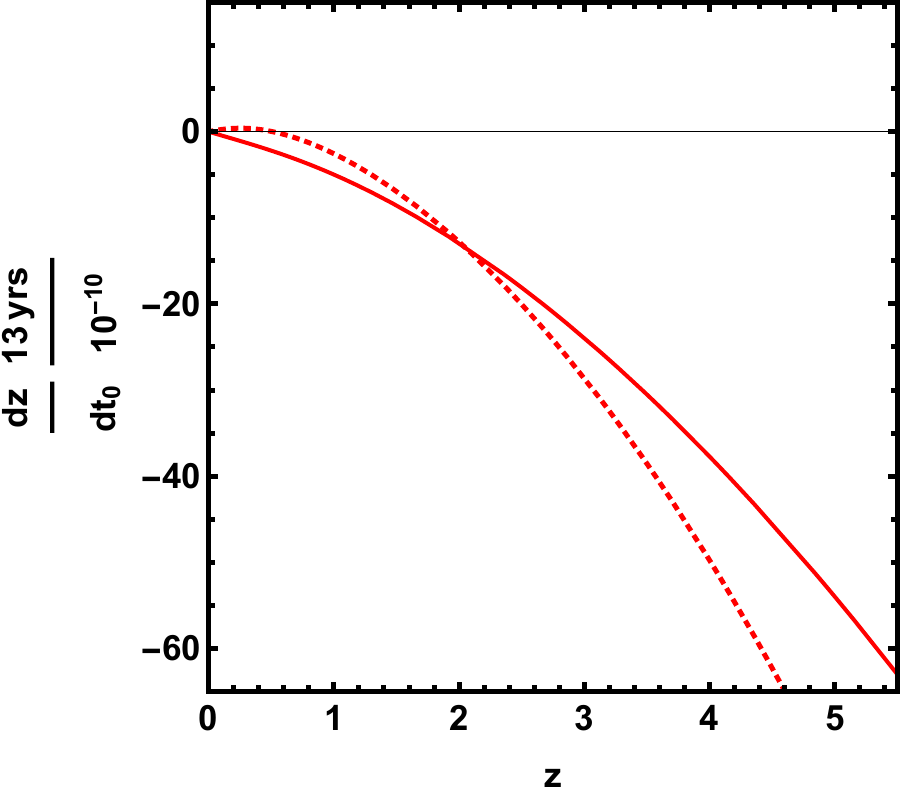}
	\includegraphics[scale=0.76]{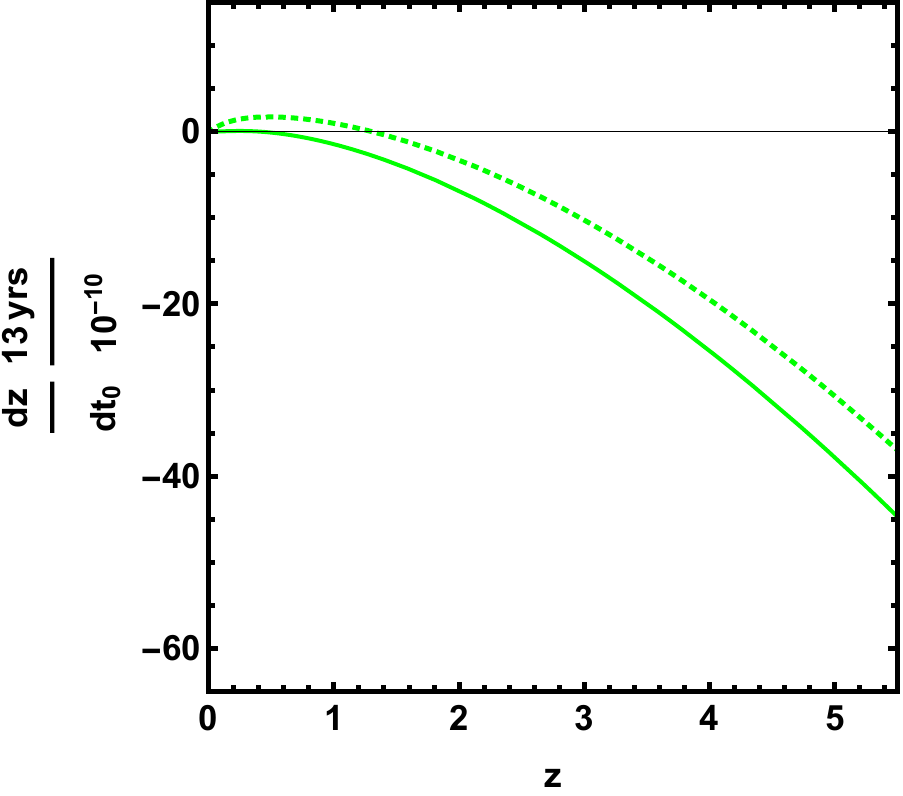}
			\caption{ Redshift drift estimated using Eq.  \ref{zd} (solid line) and using Eq. 2.28 of \cite{Dunsby_2010} (dashed line) for two different LTB scenarios: the mLTB model (left panel) and the $\Lambda$LTB model (right panel) (see \ref{sec:LTBexplicit}).}
				\label{Fig_mismatch}
		\centering
	\end{center}
\end{figure}

As was originally noted by \cite{Koksbang2016},   the  analytic  expression  for  the  redshift  drift    obtained by  \cite{Uzan2008}    (see their Eq.  9)  appears  to  differ from  the numerical results predicted by method 3,   
although  both  computational schemes correctly reproduce the standard model predictions in the FRW limit. 
This discrepancy is explicitly evident if we compare our analytical expression [cf. Eq. (\ref{zd})] with the redshift drift expression given in Eq. 2.28 of \cite{Dunsby_2010}.
We illustrate the amplitude of the mismatch  for two different LTB models, the mLTB and the $\Lambda$LTB cosmologies (described in \ref{sec:LTBexplicit}) in Fig. \ref{Fig_mismatch}.  While  \cite{Koksbang2016} suggests that the disagreement might be caused by a different definition of  the redshift drift observable,  this in reality 
results  from \cite{Uzan2008} estimating the redshift drift a simple differential, instead, as we did, as functional derivative.

\subsection{Local peculiar velocity drift}
\label{sec:motions}
 The dominant correction to the redshift drift signal arises from the time change of the peculiar velocity  of  galaxies in the time lag $\delta t_0$ between two redshift measurements. 
  Since the peculiar velocities of the various cosmic sources are, on large cosmic scales, uncorrelated, they do not contribute to the 
 signal if not in increasing its variance.  Indeed, it was shown by \cite{2008PhRvD..77b1301U},
 using linear perturbation theory of the standard model of cosmology,  that stochastic noise does  not to contribute at a level greater than $0.1\%$ for $z\leq 5$.  
  However \cite{Bel2018} showed that,  on the characteristic time scale of the redshift drift ($H_{0}$),
  the peculiar velocity of the observer (the Local Group center)  change due to the competing effect of global cosmic expansion
 and local gravitational pull of surrounding cosmic mass structures. We here quantify whether  the systematic 
drift of the observer's velocity, computed in a LTB spacetime, has any measurable effect on the redshift drift.
 
 If the observer moves ($U^i(t_0,\chi_0) \ne 0$), then the observed redshift $z^o$ of a comoving source is related to the cosmological redshift $z$
 as [cf Eq. (\ref{zdefe})]  
 \[
 1+z^o=\frac{1+z}{1+\boldsymbol{\beta}^o \cdot \boldsymbol{n}}
 \]
 where $\beta^i=\sqrt{|g_{ii}|} U^i$ is the physical (peculiar) velocity and $\boldsymbol{n}$ is the line of sight direction. 
At leading order, and setting $\beta= \alpha d\chi/dt$, 
\[ \frac{\delta z}{\delta t}=\frac{\delta z}{\delta t_0}-\dot{\beta}\]
since the photons propagates along the radial coordinate $\chi$ for an observer sitting at the center of symmetry of the LTB metric (we neglect higher order terms arising from the fact that the observer will move out of the central position due to its peculiar velocity.)
We can estimate the  additional dipole modulation of the  redshift drift signal  by computing the  radial component of the  acceleration of the central observer in a LTB metric.
From Eqs.  (\ref{offko}) and (\ref{offk1})  we obtain  that 
\[
   \dot{\beta} = -H_{\parallel}\beta \left(1-\beta^2 \right) - \frac{\alpha'}{\alpha^2}\beta^2. \]
If the LTB metric evolves according to the   EFE then $\alpha'=0$ so that  the acceleration vector is proportional to the velocity one. 
At  leading order, we  thus conclude that 
\[\dot{\beta}^o \approx H_0  \beta^o  \left ( 1+  \left .  \frac{(\log H)'}{(\log A)'} \right|_0 \right). \]
For relevant models (in which the difference between the longitudinal and the transverse Hubble function is not greater than $10\%$ 
as for example the models discussed in Sec. \ref{sec:LTBexplicit}, and in which the central  observer velocity is not greater than $\sim 700$ km/s, the standard model velocity 
of the  Local Group of galaxies with respect  to the cosmic microwave background (CMB), this  systematic effect  is not expected to exceed $1\%$ in the direction of motion of the observer. 
It is interesting to note that even if future redshift drift data were to reach a comparable level of accuracy, the dipolar  character of the contaminating signal would facilitate its 
 identification and subtraction.

\begin{figure}
	\includegraphics[scale=0.41]{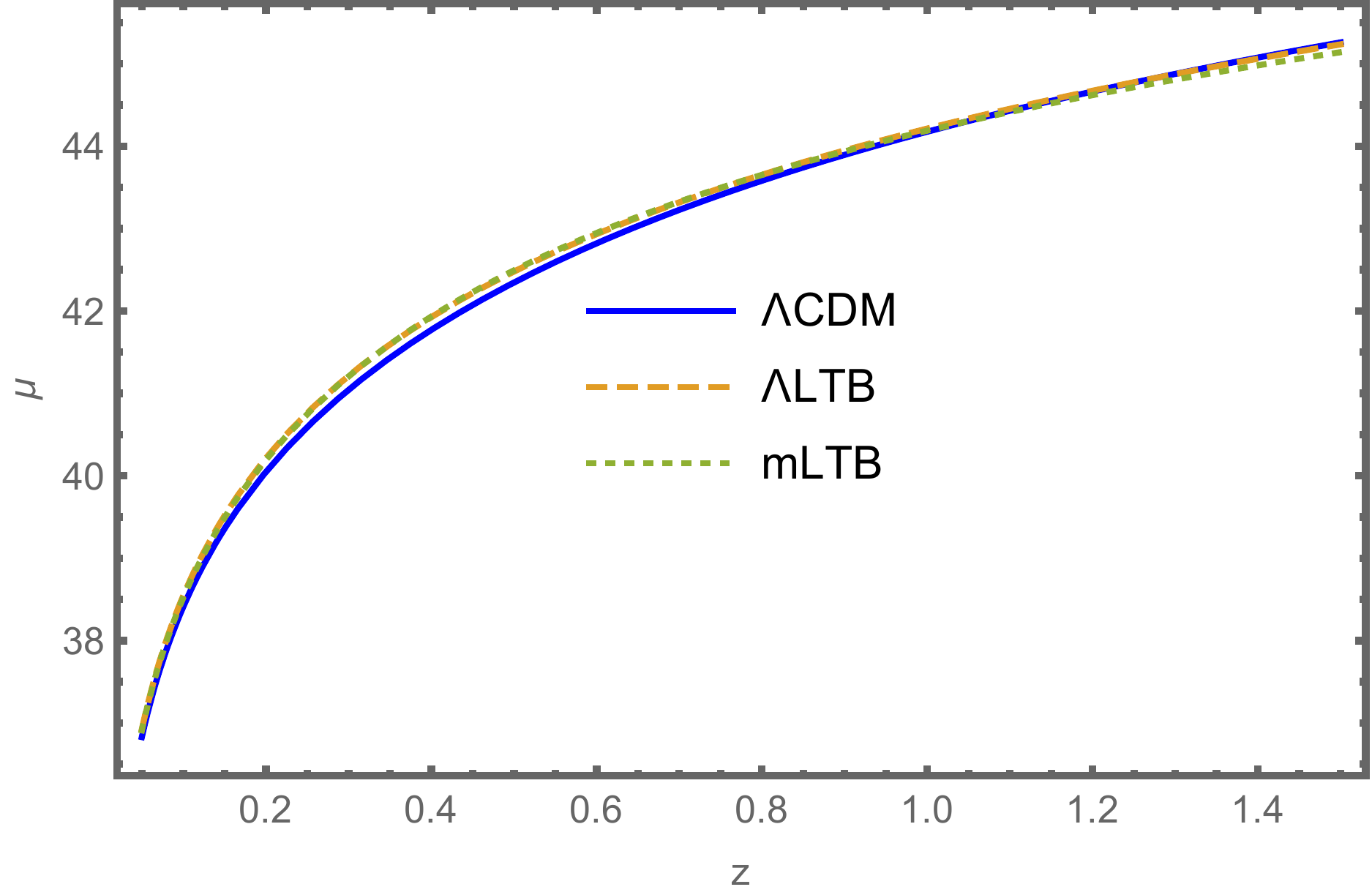}
		\includegraphics[scale=0.28]{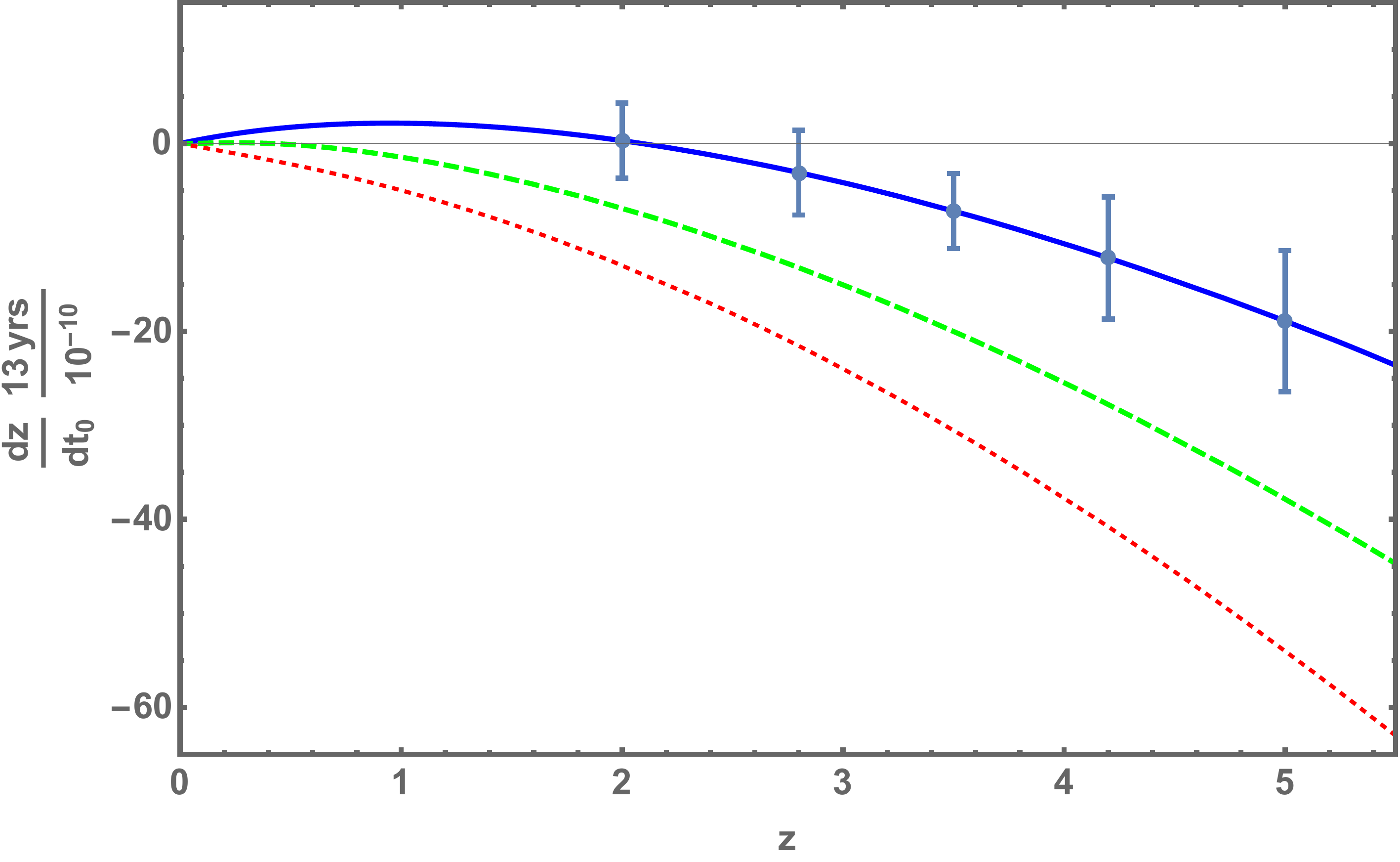}
		\caption{{\it Upper:} distance modulus calculated in  $\Lambda$CDM (solid line),  $\Lambda$LTB (dashed line)  and in $m$LTB (dotted line)  respectively. {\it Lower: } redshift drift expected in the models presented in the upper panel. The $1 \sigma$ error bars and the data range are those predicted for CODEX observations over a 13-year time span. Data points and error are taken from \cite{Quercellini2012}.}
		\label{plot:multbml}
\end{figure}

\begin{figure}
	\includegraphics[scale=0.91]{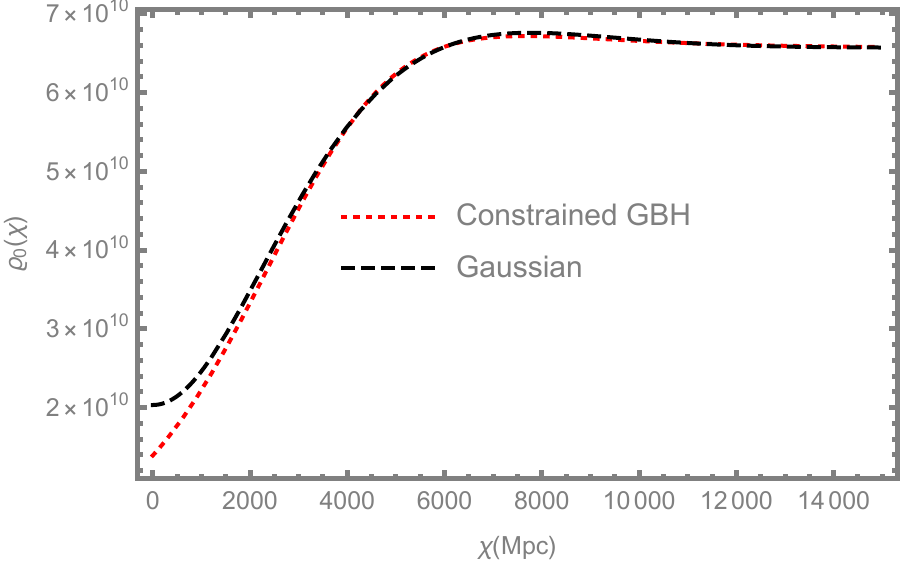}
		\includegraphics[scale=0.91]{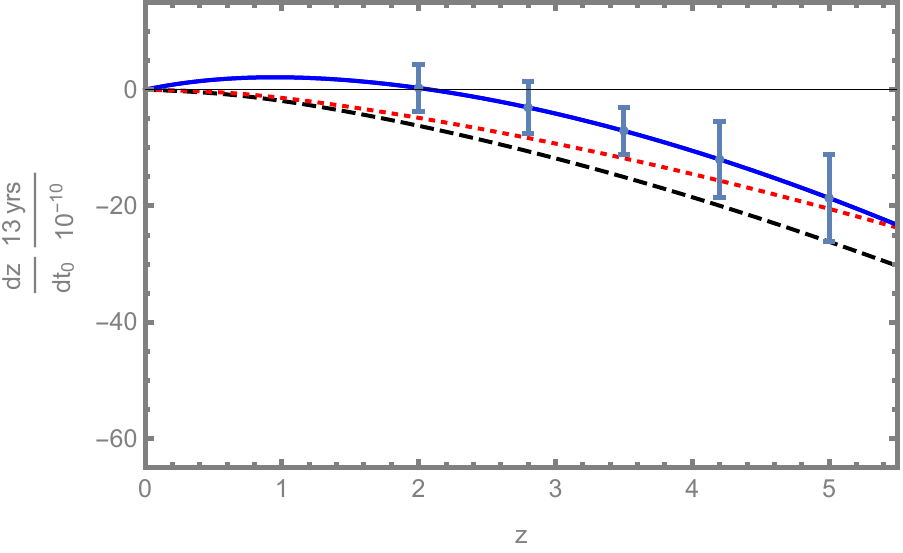}
		\caption{{\it Upper:} the spatial scaling of the present day physical matter density profile $\rho_0(\chi)$ (in units $M_{\odot} Mpc^{-3}$) for the constrained GBH (red dotted line) 
		and the Gaussian (black dashed line) void models.
		 {\it Lower: } comparison of the redshift drift expected in the standard $\Lambda$CDM scenario (blue solid line) and in the above LTB models  which best fit SNIa, BAO and CMB data. Data points and 1 $\sigma$ error are the same as in Fig. 3.}
		\label{plot:fig4}
\end{figure}

 \section{Application  with explicit LTB models} \label{sec:LTBexplicit}

In this section, we review some popular LTB models published in the literature that claim to reproduce the predictions of the $\Lambda$CDM model.
We premise that we are not interested in their physical feasibility or cosmological soundness, but in their practical utility as toy models in order to highlight the potential of the redshift drift observable  in resolving some of the degeneracies that plague LTB cosmologies when compared to observational data. 

The first set of models are those that were shown by \cite{Enqvist2007} to reproduce the  Hubble diagram of supernovae. 
The first   model  (we call it mLTB) describes a LTB  universe which is comprised only  of matter ($\Omega_{m0}(\chi)=1$), while the second one ($\Lambda$LTB) incorporates the contribution of the cosmological constant  $\Lambda$ and satisfies to the flatness constraint $\Omega_m(\chi)+\Omega_{\Lambda}(\chi)=1$. The rationale for choosing them  is that  the spacetime evolution of the  scale factor $A(t,\chi)$ can be computed analytically, making the results more transparent. 
By integrating Eq. \eqref{ltb1} for $\Omega_m+\Omega_\Lambda=1$ we obtain  the following expression for the scale factor:
\begin{align}
    \frac{A}{A_0} & =\left[\cosh\left(\frac{3}{2}\sqrt{1-\Omega_{m,0}}H_0(t-t_0) \right)\right.\\
    &+\left. \frac{1}{\sqrt{1-\Omega_{m,0}}}\sinh\left(\frac{3}{2}\sqrt{1-\Omega_{m,0}}H_0(t-t_0) \right) \right]^{2/3} \nonumber
\end{align}
which reduces to $A/A_0=[1+3/2H_0(\chi)(t-t_0)]^{2/3}$ in the case of the mLTB model. 

Both    models are described by  the same present day transverse expansion rate profile 
\begin{equation}
    H_0(\chi) = H_0(\infty) + \Delta H e^{-\chi/\chi_0}
    \label{Hochi}
\end{equation}
with parameters fine tuned to fit the Hubble diagram of SNIa data (see Fig. \ref{plot:multbml}). Specifically, the mLTB model is characterised by 
$H(\infty)=48.7$ km/s/Mpc, $\Delta H = 16.8$ km/s/Mpc and $\chi_0 = 1400$ Mpc, while the $\Lambda$LTB scenario has best fitting parameters  $H_0(\infty)=58$ km/s/Mpc, $\Delta H = 8$ km/s/Mpc,  $\chi_0 = 600$ Mpc and $\Omega_{\Lambda 0}(0) = 0.33$. 

The second set includes more sophisticated models that are tuned to agree with multiple observational probes, not just the Hubble diagram of SNIa. They are also meant    to satisfy theoretical stability constraints.  For example the simple inhomogeneous models mLTB and $\Lambda$LTB  share the assumption that the  the universe  has  an inhomogeneous big bang, {\it i.e.}  it came into being at different times at different positions $\chi$. However, variations in the cosmic age function can be related to the decaying modes in the theory of linear perturbations \cite{SilkJ} which,  in turn,  would imply the existence of large fluctuations at remote epochs, in contradiction with the  remarkable homogeneity of the CMB spectrum. On the contrary , it is much more constructive, from a physical point of view, to assume that spherical inhomogeneities develop and grow  over time, due to gravitational instabilities, in an otherwise uniform universe. Therefore, we would like to consider scenarios in which a spherically symmetric inhomogeneity is asymptotically embedded in the uniform FRW spacetime, i.e. converge to the standard metric of the universe at large distances and at early times. This is achieved by considering 
models in which the  present day reduced matter density parameter is
\begin{equation}
\Omega_{m0}(\chi)=\Omega_{m 0}^s \left(1+\delta_{\Omega}\right)
\end{equation}
where $\Omega_{m0}^s$ is the reduced density of the embedding standard flat $\Lambda$CDM model.
We thus extend the set to include the constrained model of  \cite{Bellido2008a} (hereafter constrained GBH)
\begin{equation}
\delta_{\Omega}=\delta_v  \frac{1-\tanh \left(\frac{\chi -\chi_{s}}{2 \Delta \chi_s} \right)}{1+\tanh \left(\frac{\chi_s}{2 \Delta \chi_s}\right)}
\label{deltaom}
\end{equation}
where $\delta_v$ is the relative  difference  between the reduced density parameter at the  center of the void and the asymptotic FRW value 
 $\chi_s$ is the typical size of the void,  and the parameter $\Delta \chi_s$ controls the steepness of the transition between the interior and exterior of the underdensity. 
When implemented with carefully tuned parameters, this model  is in agreement with the Hubble diagram of SNIa, the Baryon Acoustic Oscillation (BAO), the CMB and also the age of old high redshift objects \cite{age}. For the purposes of the present analysis, we adopt  the best parameters fitted by  \cite{Vargas}
( $\Omega_{m0}^s=1, \delta_v=-0.88,  \chi_s=3.72$ Gpc,  $\Delta \chi_s=1.68$ Gpc). 

Additionally, we also also consider the Gaussian void model \cite{Clifton_2008},
\begin{equation}
\delta_{\Omega}=\delta_v e^{- \left(\frac{\chi}{\chi_s}\right)^2}
 \label{deltaom}
\end{equation}
where the parameters have the same physical interpretation as in the previous case.
We implement this void profile by setting  $\Omega_{m0}^s=1, \delta_v= -0.81,  \chi_s=5.04$ Gpc, 
which, according to  \cite{Vargas}, provide an equally satisfactory fit to the SNIa, BAO and CMB data. 
For both the constrained GBH and the Gaussian LTB models  the spatial hypersurface at the big bang does not depend on
the radial coordinate $\chi$, meaning that the radial scaling of the present day transverse expansion rate is fixed  by the relation
\[
H_0(\chi)=\frac{H_0}{1-\Omega_{m0}}
\left[ 1-\frac{\Omega_{m0}}{\sqrt{1-  \Omega_{m0}}}   \sinh^{-1}\sqrt{\frac{1- \Omega_{m0}}{\Omega_{m0}}}\right]
\]
where  $H_0$ is chose so that the age of the universe is $13.8$ Gyrs.
For both these matter+curvature models the time and spatial evolution  of the transverse scale factor is  given by the parametric equations 
\begin{eqnarray}
A(\eta, \chi) & = & \chi \frac{\Omega_{m0}}{1-\Omega_{m0}}\left( \cosh \eta-1\right)  \nonumber \\
t( \eta, \chi) & = & \frac{1}{H_{0} (\chi)} \frac{\Omega_{m0}}{2(1-\Omega_{m0})^{3/2}}\left( \sinh \eta-\eta\right).  \nonumber
\end{eqnarray}
The resulting void density profiles, obtained by using eqs. \ref{md} and \ref{eq:rhoti}), 
are shown in the upper panel of Fig. \ref{plot:fig4}.

\begin{figure}
	\begin{center}
		\includegraphics[scale=0.45]{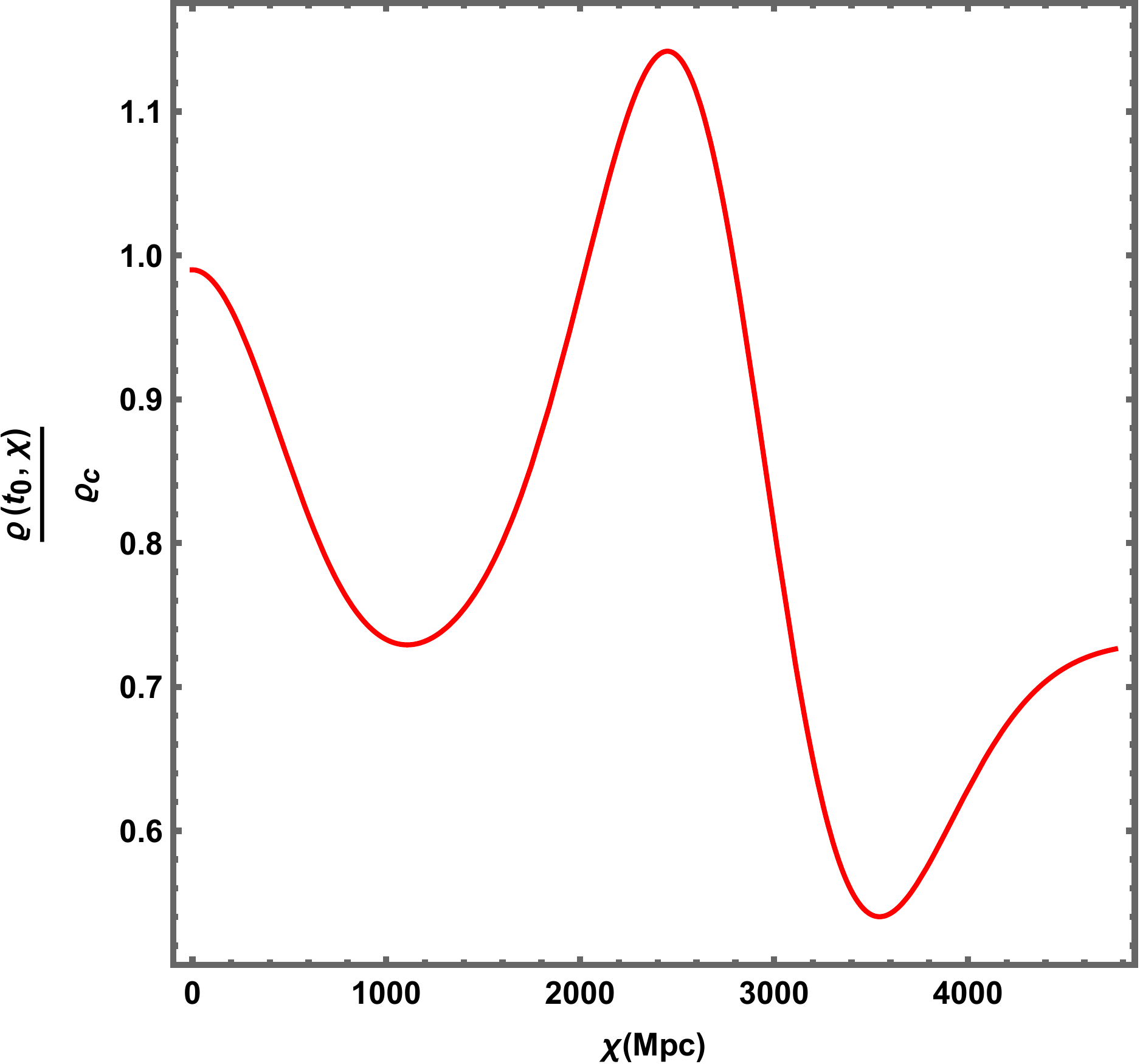}
		\includegraphics[scale=0.45]{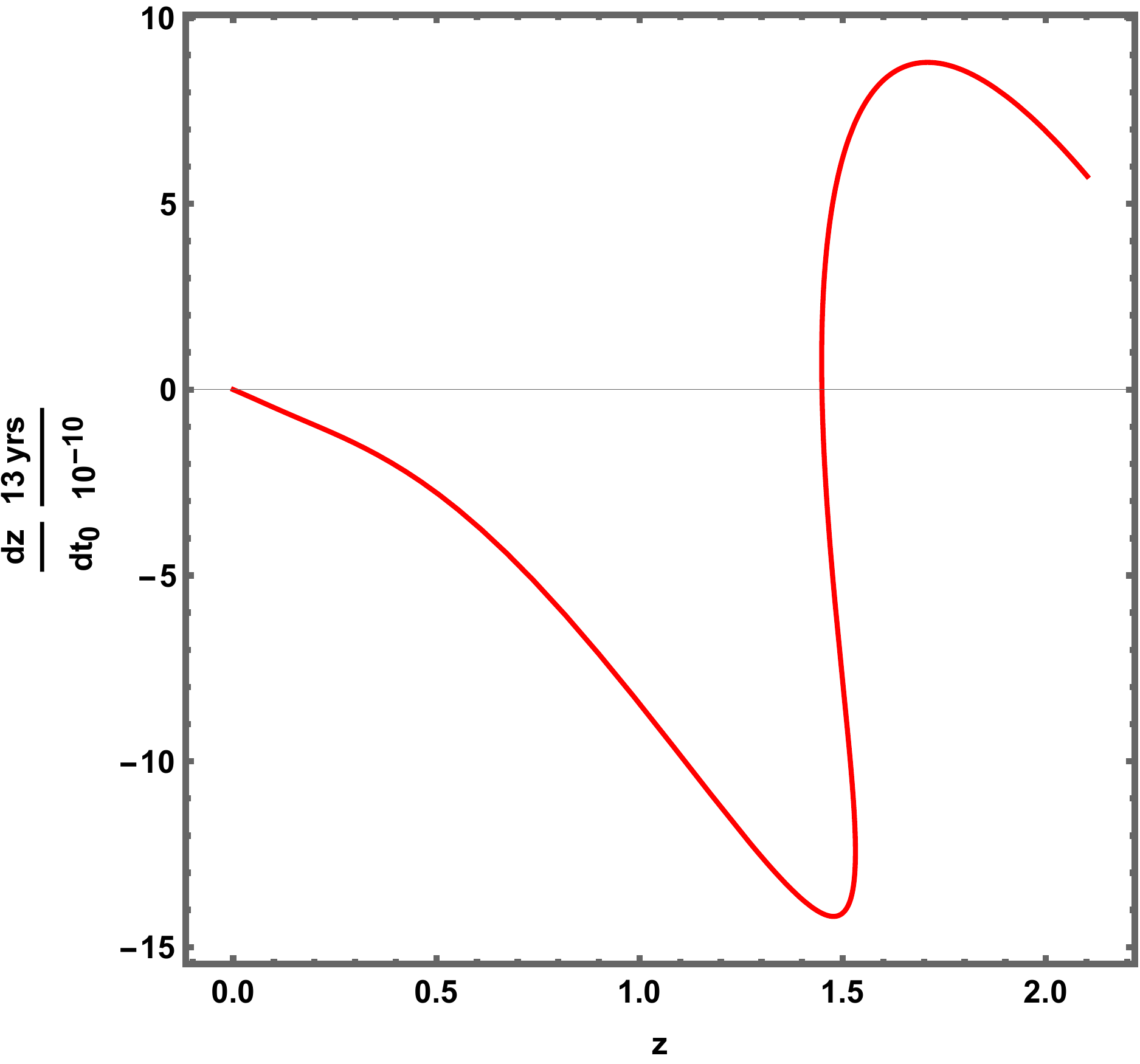}
		\caption{Redshift drift in a cosmological model where the redshift is a non  monotonic function of the radial coordinate $\chi$. {\it Upper:} matter density profile. {\it Lower:} redshift drift.}
		\label{plot:omegam}
		\centering
	\end{center}
\end{figure}

\begin{figure}
	\begin{center}
	        \includegraphics[scale=0.32]{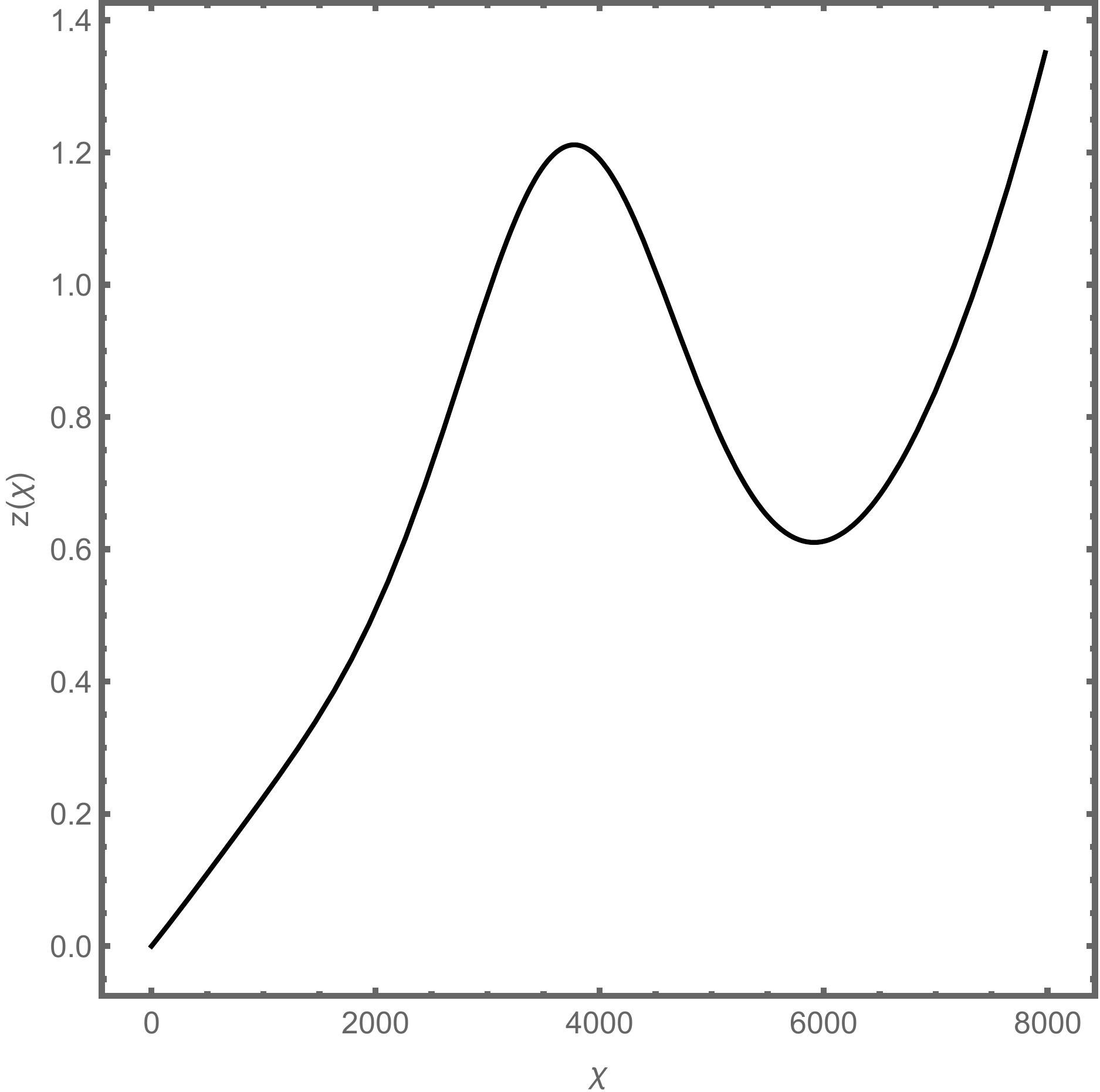}
		\includegraphics[scale=0.32]{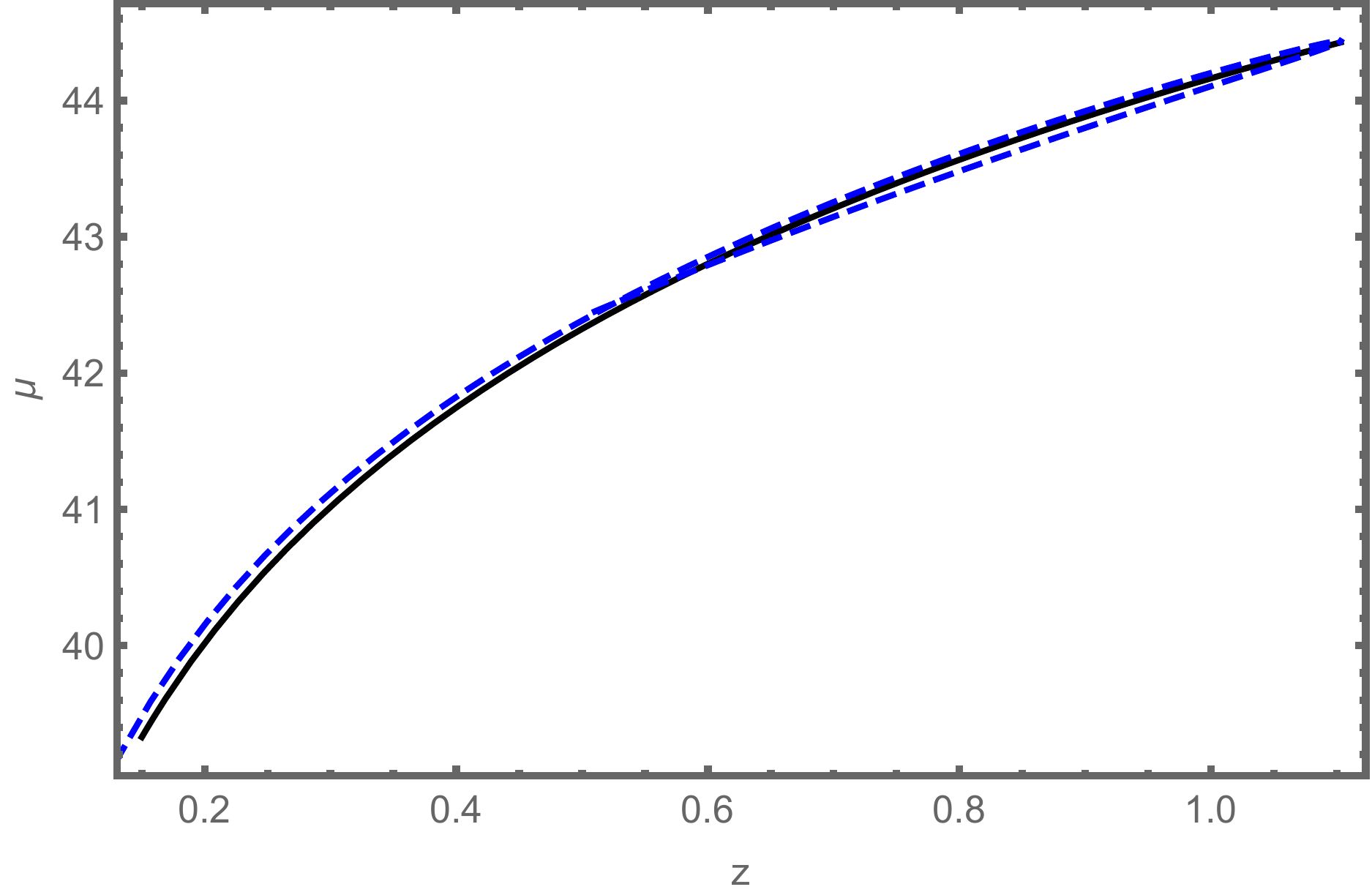}
		\includegraphics[scale=0.32]{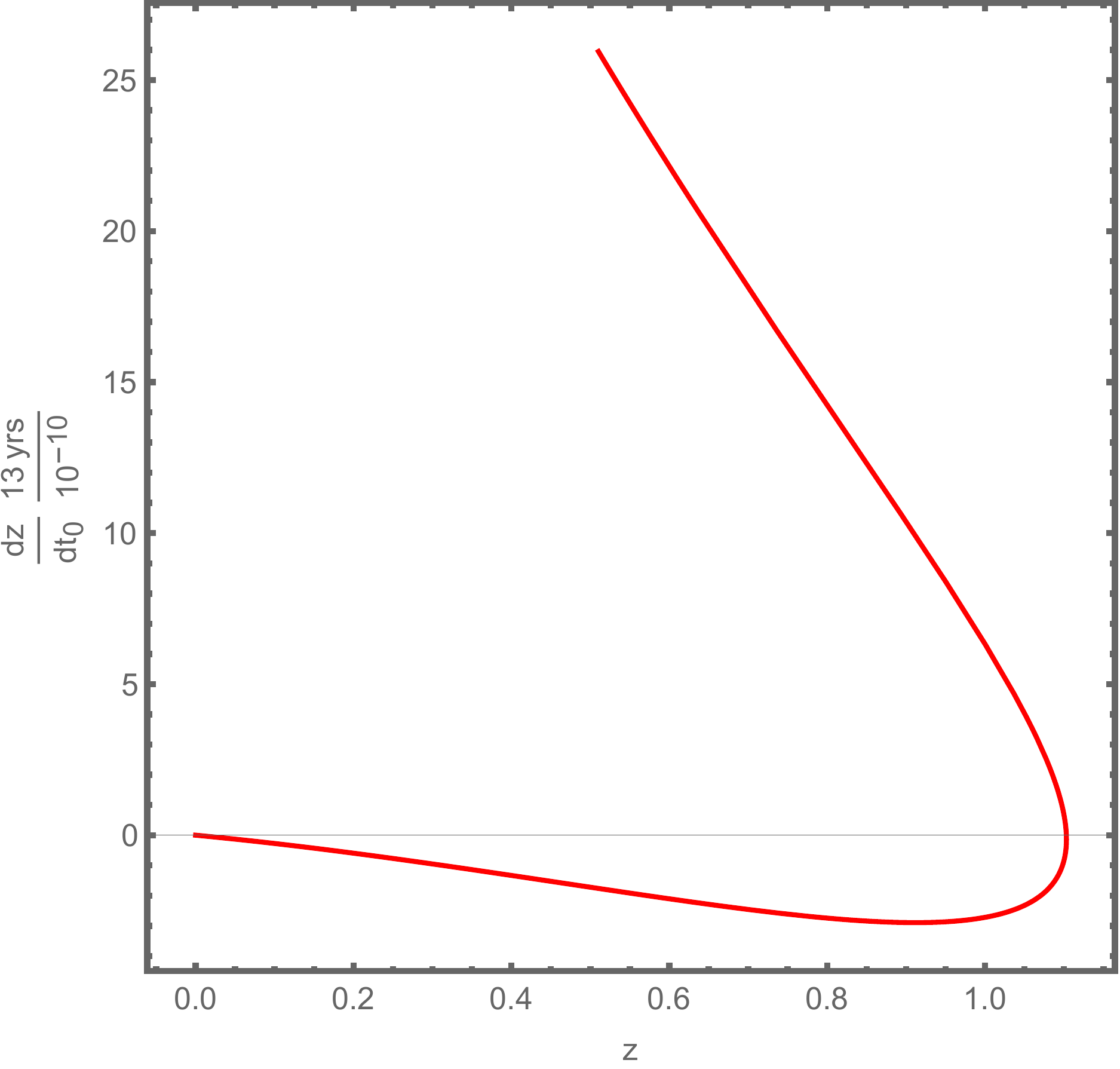}
		\caption{{\it Upper panel:} redshift $z$ as a function of the radial  coordinate $\chi$ along the null geodesic of a photon in  a LTB model in which   $\chi(z)$ is not a single valued function. 
		{\it Central panel:}  the Hubble diagram is plotted as a function of redshift for the LTB model shown above. In this particular model, photons emitted at two different epochs  have the same redshift and the same transverse scale factor $A(t, \chi(t))$ and thus their Hubble diagram is degenerate and indistinguishable from that of the standard  $\Lambda$CDM model. {\it Bottom panel:} redshift drift as a function of redshift for the same LTB model shown in the upper panel.}
		\label{plot:bivalue}
		\centering
	\end{center}
\end{figure}

\subsection{Discussion}
The luminosity distance  of an object at redshift $z$ is given by (see for example  \cite{Ellis:2009aa})
\begin{equation}
    d_L(z)=(1+z)^2 A(\chi(z),t(z))
\end{equation}
where $\chi(z)$ and $t(z)$ are computed along the null geodesics of the photon  using equations \eqref{dzdt} and \eqref{dzdchi}. 
The  resulting  distance modulus $\mu$ as a function of the redshift 
for the $\Lambda$CDM model and the mLTB and $\Lambda$LTB models 
is shown in the upper panel of Fig. \ref{plot:multbml}.  
By construction they do not differ appreciably. Indeed,  
because of the additional functional degrees of freedom induced by the spatial dependence of the transverse expansion rate, the Hubble diagrams of many physically distinctive  inhomogeneous models are degenerate.
This is mostly the consequence of this observable  being an integral quantity of the redshift. If instead of the distance to cosmic objects we consider  the time change of their redshift, we can restore some degree of predictability. This is shown on the bottom panel of Fig. \ref{plot:multbml}  where the redshift drift for these very same three cosmological models is 
plotted as a function of redshift. LTB models whose magnitude-redshift predictions are degenerate with the standard Hubble diagram can be easily distinguished because they predict very different behaviors for the redshift drift. 
By inspecting   Fig. \ref{plot:multbml}  one can also appreciate how the data expected from future observational projects (the CODEX survey in this case) have the power necessary to discriminate not only the standard model from more exotic LTB-type scenarios, but also between various LTB models that would otherwise be indistinguishable on the basis of their Hubble diagram.

There are also LTB models that are compatible with a wider range of data, particularly the BAO and CMB measurements, which, in addition to predicting the same luminosity distance, are also characterized by void density profiles that are observationally indistinguishable. 
This is the case, for example, of the constrained GBH model and the Gaussian model shown in the top panel of Fig. \ref{plot:fig4}. Interestingly, also these models 
predict redshift drift signals that are significantly different not only from the expectations of the standard  $\Lambda$CDM model, but also from each other. 

One can better appreciate the specificity of this phenomenon by contrasting it to what happens in the standard model. 
Even in a homogeneous and isotropic spatial background, two different models of dark energy (one for example in which the acceleration of the scale factor of the metric is contributed by the cosmological constant and another in which it originates from the `negative' pressure exerted by a scalar field) can be fine-tuned to reproduce the same Hubble diagram  of SNIa (see for example \cite{Perenon}). However,  the redshift drift cannot disentangle these two different physical models since 
the predicted amplitude  will also be degenerate.   

The ability of the redshift drift probe in discriminating different inhomogeneous LTB models relies entirely on two essential features. First, although
LTB models have a large number of functional degrees of freedom, at most only a subset of structural functions characterizing different void models can be indistinguishable. In particular, the various models considered in our analysis all exhibit the same spatial and temporal dependence of the transverse expansion rate $H$, although the scaling of the longitudinal expansion rate $H_{\parallel}$ turns out to be very different and model dependent. 
For example, the maximum amplitude of the ratio $H/H_{\parallel}$, which occurs at about $\chi \sim \chi_s$, is significantly larger $(\sim 20\%)$ in the constrained GBH model than in the Gaussian model. Consequently, observables such as redshift drift, which are sensitive to multiple structural parameters (in particular to $H_{\parallel}(t, \chi), H(0, t)$, and $\alpha (t, \chi)$ as can be seen in Eq. \ref{zd}) and not only to the transverse expansion rate, as in the case of the luminosity distance, can be effectively used to singularize LTB models. In this sense, the Hubble diagram, like any observable sensitive to a limited set of LTB structural functions, is a poor diagnostic tool.

In addition, the functional nature of the redshift drift in spherically symmetric spaces is a key feature that facilitates model discrimination.
This observable  does not depend only on the redshift of the emitting source, as in the standard model of cosmology,  but also on the geodesics taken by the photons to reach the observer. The path integral that goes into its definition (cf. (\ref{zd})) makes the redshift drift in spherically symmetric spaces a nonlocal and explicitly model-dependent quantity.
 For example, fine tuning the scaling of $H$ in different LTB models to reproduce the luminosity distance  comes at the price of making the geodesic path of photon to vary from model to model. Observable as the redshift drift,  being sensitive to the entire geodetic history of the signal, are therefore optimal tools for distinguishing various inhomogeneous patterns. 

The non-local nature of the redshift drift  in LTB cosmologies is responsible of some interesting phenomenology. 
The behaviour of the redshift drift at   low $z$ follows from taking the lowest order terms of the series expansion of Eq. (\ref{zd}) around $t=t_0$
\[\frac{\delta z}{\delta t_0}\approx  -\left .\frac{H'_{\parallel}}{\dot{\alpha}}\right |_{0}\delta t_0+H_{\parallel}(0,t_0)z-
\dot{H}_{\parallel}(0,t_0)\delta t_0 - H'(0,t_0)\delta \chi \]
which reduces to 
\begin{align}
\frac{\delta z}{\delta t_0} \approx \frac{\ddot{\alpha_0}}{\dot{\alpha_0}}z= -\frac{1}{2} H_0(0) z \left( \Omega_{m0}(0) - 2 \Omega_{\Lambda_0}(0) \right).
\end{align}
According to the above equation,  the redshift of a low-$z$ source decreases in time  in a purely matter-dominated LTB model. In other terms, if dust is the only source of gravity,   the   redshift drift cannot be positive for objects close to the center of symmetry. In this respect,  the matter-dominated LTB models behave  as a  homogeneous and isotropic universe filled with dust. However, contrary to standard model results, the redshift drift might eventually becomes positive at higher redshifts. This  effect is shown in 
Fig. \ref{plot:omegam} together with the peculiar radial matter  density profile that generates it. The phenomenon is induced by the 
 non-monotonic scaling of $\alpha(\chi)$ which causes the functions $H_{\parallel}$ and 
 $\partial_{\chi}H_{\parallel}$ 
 to change  sign along the photon geodesc. In this specific case $H_{\parallel}$ becomes progressively more negative as the radial coordinate of the source increase, forcing the sign of the redshift drift to turn positive (see Eq. \ref{zd}).

Other unconventional effects characterises  the redshift drift in LTB cosmologies. 
One can engineer the shape of $H_0(\chi)$ such that redshift 
itself is no longer a monotonic function of the spacetime coordinates $t$ and $\chi$, 
or, which is equivalent,  the time and radial coordinates are no longer  single valued functions of the redshift  along the null-geodesic path. 
One can exploit this functional degree of   freedom associated to $H_0(\chi)$,  for example, to design models where the redshift of very distant objects 
turn negative. More intriguingly, one can also devise physical models,  i.e. LTB  models with  a strictly  non negative matter density function $\rho_m$,  where  objects at different radial coordinates $\chi$ (thus emitting the photons at different epochs) have the same redshift and, thus, the same radial scale factor $A(t(z), \chi(z)$, or, equivalently, the same distance modulus $\mu(z)$. 
This peculiar phenomenon  is shown in Fig. \ref{plot:bivalue}, where the Hubble diagram is plotted as a function of redshift.  
As the photons coming from these sources are not following the same geodesics path, the Hubble diagram  degeneracy 
is broken  when the redshift drift is analysed. Indeed, since the redshift drift is a non-local, path dependent  function, 
two different values of the redshift drift can 
correspond to the same redshift value. In this context,  means that the density matter is strictly non negative function.

\section{Conclusion}\label{sec:conclusion}
Providing evidences of cosmological changes on human time-scales \cite{Quercellini2009,Bel2018,Heinesen_2021_b} is challenging. While the physics is transparent and fascinating, the signal is characteristically weak. Nevertheless, measurements of the redshift drift are certainly within the reach of large telescopes in the next 20 years \cite{Bolejko2019}. It is thus worthwhile exploring the possibilities for  using this observable to understand the  cosmological impact of  inhomogeneities in the large-scale structure of the universe. 

In this context, our goal was to provide an analytical formula for calculating the redshift drift in spherically symmetric spaces. As a result, we have highlighted
the functional character of this probe, a quantity that depends not only on the time instants at which photons from a source are emitted and received, but also on the 
null-geodesics traveled by the photons to reach the observer. 

The non-local nature of the redshift drift in spherically symmetric spaces is at the origin of some peculiar  phenomena. Arguably, the most interesting is the possibility of disentangling inhomogeneous models predicting the same Hubble diagram. 
Since the LTB spacetime has fewer symmetries than the standard metric of the universe,  the
distance-redshift relation of the standard $\Lambda$CDM  model can  be reproduced  with arbitrary precision using, for example, the freedom in the specification  
of the transverse expansion rate function $H_0(\chi)$.  Even more maliciously,  various inhomogeneous models, characterized by widely different structural parameters,  predict 
distance moduli so similar that they resist the resolving power of data. 

We have shown that 
the redshift drift has the potential  to break this degeneracies and restore some predictive power to LTB models. 
This is important if we consider that the linear perturbation theory of LTB cosmological models, because of the technicalities it implies, is still in its infancy and that observable features of the large-scale structure of the universe, such as peculiar velocities, growth rates or  power spectra matter density fluctuations etc.,  cannot be effectively used to resolve degeneracies appearing in the background sector of the theory. 

One possible way to progress further and complete the study would be to evaluate 
 the consequences of moving the observer's position away from the center of symmetry  and 
 quantify the dependence of the redshift drift on the specific off-centric location. 
 We plan to expand on these issues in a forthcoming  paper.

 \vskip 1.truecm

 \noindent {\bf Acknowledgements}.
We would like to thank   Julien Bel and Federico Piazza   for useful discussions. 
This work was partially supported  by the \textit{Institut  Physique de l'Univers} (IPHU Grant No. 013/2020)
 and by the Programme National GRAM of CNRS/INSU with INP and IN2P3 co-funded by CNES.

\bigskip
\bigskip

\appendix

\section{Evolution of the scale factors of the LTB metric}

The Einstein field equations  for a LTB spacetime containing a non-relativistic  perfect fluid, {\it i.e.} dust with an equation of state $(p=0)$, and a 
non-dilutive dark energy component  $\rho_{\Lambda}=const$ are 

\begin{equation} \label{fried00}
    \frac{1}{A^2}+\left(\frac{\dot{A}}{A}\right)^2+2\frac{A'\alpha'}{A\alpha^3}-2\frac{A''}{A\alpha^2}+2\frac{\dot{\alpha}\dot{A}}{A\alpha}-\left(\frac{A'}{A\alpha}\right)^2=8\pi G \left(\rho _m + \rho _\Lambda\right)
\end{equation}

\begin{equation} \label{fried01}
    \dot{A}' = A' \frac{\dot{\alpha}}{\alpha}
\end{equation}

\begin{equation} \label{fried02}
    \frac{1}{A^2}+2\frac{\ddot{A}}{A}+ \left( \frac{\dot{A}}{A} \right) ^2-\left( \frac{A'}{A\alpha} \right)^2 =8\pi G \rho_\Lambda
\end{equation}

\begin{equation} \label{fried03}
    \frac{\ddot{A}}{A}-\frac{A''}{A\alpha^2}+\frac{\dot{A}\dot{\alpha}}{A\alpha}+\frac{A'\alpha'}{A\alpha^3}+\frac{\ddot{\alpha}}{\alpha} = 8 \pi G \rho_\Lambda.
\end{equation}

Eq. \eqref{fried01} can be integrated with respect to $t$ resulting in
\begin{equation}\label{relaa}
    \alpha(\chi,t)=C(\chi)A'(\chi,t)
\end{equation}
where $C$ is a function of $\chi$ only which, being arbitrary, we redefine as $C(\chi)=1/\sqrt{1-k(\chi)}$.  
Once Eq. \ref{relaa} is inserted in  \eqref{fried00}, \eqref{fried02} and Eq. \eqref{fried03}, the system reduces to  the two independent equations
\begin{equation} \label{fried1}
    \frac{\dot{A}^2+k}{A^2}+\frac{2\dot{A}\dot{A}'+k'}{AA'} = 8\pi G (\rho_m + \rho_\Lambda)
\end{equation}
and
\begin{equation} \label{fried2}
    \frac{\dot{A}^2+2A\ddot{A}+k}{A^2} = 8\pi G \rho_\Lambda.
\end{equation}
We can further simplify  by multiplying Eq. \eqref{fried2} by $A^2\dot{A}$ and integrating it over time. We  find
\begin{equation}
    A\dot{A}^2 = \frac{8\pi G}{3} \rho_\Lambda A^3 - kA + \frac{8\pi G}{3} A^3 \Tilde{\rho} 
    \label{intstep}
\end{equation}
and, thus,  
\begin{equation} \label{motionltb1}
    \left(\frac{\dot{A}}{A} \right)^2 +\frac{k}{A^2}= \frac{8\pi G}{3} \left ( \tilde{\rho} +\rho_{\Lambda}\right)
\end{equation}
where $\Tilde{\rho}(t,\chi) $ is an arbitrary  integration function such that the product $A^3\Tilde{\rho}$ depends only on the radial 
coordinate $\chi$.  Its relation to the physical matter density $\rho_m$, the quantity that appears in the stress energy tensor, is obtained as follows. 
First we derive  \ref{intstep} with respect to the radial coordinate and  replace the result into Eq. \eqref{fried1} to obtain

\begin{equation}\label{densf}
 (A^3\Tilde{\rho})' =3  \rho_m A^2A'.
\end{equation}
which, upon integration, gives
\begin{equation}
    \Tilde{\rho}(t, \chi)= 3 \frac{\int_0^\chi \rho_m A^2A'd\chi}{A^3}
\label{rhot2}
\end{equation}
where we used the fact  that $\Tilde{\rho}(0,t)A^3(0,t)=0$.
The inverse relationship can be recovered by performing the differentiation in (\ref{densf})
\begin{equation}
    \rho_m
     = \Tilde{\rho} + \Tilde{\rho}' \frac{A}{3A'}.
\end{equation}

The flat average density of matter $\Tilde{\rho}$ allows us to rewrite the equations of motion of the scale factor $A(t,\chi)$ of the LTB metric in a way similar to those governing the evolution of the scale factor $a(t)$ of the FRW metric in the standard model of cosmology, {\it i.e.} in  a Friedmann-like form. Indeed,  by deriving  with respect to time Eq.  (\ref{motionltb1}) as well as  the time-independent product  $A^3 \Tilde{\rho}$ we get
\begin{equation}\label{acceltb}
\frac{\ddot{A}}{A}=-\frac{4 \pi G}{3} \left( \tilde{\rho}-2\rho_{\Lambda}\right)
\end{equation}
and
\begin{equation}\label{contltb}
    \dot{\Tilde{\rho}} + 3\frac{\dot{A}}{A}\Tilde{\rho}=0
\end{equation}
{\it i.e.} the acceleration and continuity equations respectively.  

The acceleration of the longitudinal scale factor is 
\begin{eqnarray}\label{acceaa}
\frac{\ddot{\alpha}}{\alpha}=\frac{\ddot{A}^{\prime}}{A^{\prime}} & = & -\frac{4}{3}\pi G\left[  \frac{ (\tilde{\rho} A)'}{A'} -2\rho_{\Lambda}\right] \nonumber \\
&=& \frac{\ddot{A}}{A}-\frac{4}{3}\pi G \left[ \frac{(\tilde{\rho} A)'}{A'} - \Tilde{\rho}  \right] 
\end{eqnarray}
which, in the FRW limit (no radial  inhomogeneities), gives $\ddot{\alpha}/\alpha=\ddot{A}/{A}$. 

The longitudinal and transverse expansion rates are related as 
\begin{equation}
H_{\parallel}=H+\frac{A}{A'}H'=H+\frac{A^2}{2\dot{A}A'}\left[\frac{8\pi G}{3}\tilde{\rho}'- \left( \frac{k}{A^2} \right)'\right]
\end{equation}
and, at the center of symmetry ($\chi=0$),  they coincide $H_{\parallel}(0, t)=H(0, t)$.

\section{Time dilation formula in a LTB cosmology}
\label{sec:appA}

Although the time dilation formula (\ref{dilations}) can be taken as the very definition of what  redshift is, i.e. the relative change in the proper frequency of a signal 
as measured at emission and at reception \cite{Ellis:2009aa}, it is  instructive to see how this result can be explicitly derived in the framework of the LTB spacetime. 
In doing so, we will have the opportunity to restate the formula we obtained for redshift (cf. Eq. (\ref{zf})) by means of an alternative argument. 

Consider two light rays emitted at $t$ and $t+\delta t$  by the very same source. Suppose they are received by the same detector at time $t_0$ and $t_0+dt_0$.
The geodesic equation of the first ray is 
\[t_1=t(\chi)\]
while that of the second ray is 
\[t_2=t(\chi)+\delta t(\chi).\]

It is clear that they are both solutions of the geodesic equation with the boundary conditions $t_1(\chi)=t(\chi)$,  $t_1(0)=t_0$,  $\delta t(\chi)=\delta t$, $\delta t(0)=\delta t_0$
and $t_2(0)=t_0+\delta t_0$. By inserting them into  (\ref{geode}) one gets
\begin{eqnarray}
\frac{dt(\chi)}{d\chi} & = & -\alpha(\chi, t(\chi)) \\
\frac{d\delta t(\chi)}{d\chi} & = & 
-\delta t(\chi) \dot{\alpha}(\chi, t(\chi)). 
\end{eqnarray}

The second equation describes the time dilation and can be straightforwardly solved  to give 
\begin{equation}
\frac{\delta t_0}{\delta t }=e^{\int_0^{\chi}\dot{\alpha}(\chi, t(\chi))d\chi}
\end{equation}
where the integral is calculated along the null geodesic of the photon.  
Since Eq. (\ref{zf}) can be rewritten as 
\begin{equation}
1+z=e^{\int_0^{\chi}\dot{\alpha}(\chi, t(\chi))d\chi}
\end{equation}
we deduce that 
\begin{equation}
1+z=\frac{\delta t_0 }{\delta t}.
\end{equation}

\bigskip

\section{Numerical redshift drift}
\label{sec:appB}

The numerical algorithm for calculating the redshift drift (referenced as {\it method 3} in Sec. \ref{sec:bias}) is here briefly discussed. 
The null-geodesics of a photon (labelled by the index 1) emitted at time $t$ and received at time $t_0$ by two galactic sources   is parametrized by the equations
\begin{align} 
z_1 & =z(\chi)  \nonumber \\
t_1 & =t(\chi)  \nonumber 
\end{align}
as a function of the radial comoving coordinate $\chi$. 
A second photon,   emitted and received  by the same sources  at the later times $\delta t$ and $\delta t_0$,  will move along the path
\begin{align} 
z_2 & =z(\chi)+\delta z(\chi) \label{z2} \\
t_2 & =t(\chi) +\delta t (\chi) \label{t2}
\end{align}
where the following   boundary conditions are assumed
\begin{align}
t(0) & = t_0, \nonumber  \\
z(0) & = 0,  \nonumber  \\
\delta z(0) & = 0, \nonumber  \\
\delta t(0) & =\delta t_0. \nonumber 
\end{align}

By substituting the photon path $(\ref{z2})$ and $(\ref{t2})$  into the equations for the redshift (\ref{dzdt}) and (\ref{dzdchi}) one gets, at leading order,  
\begin{align}
\frac{d \delta z}{d \chi} & =\dot{\alpha}(\chi, t(\chi)) \delta z+(1+z)\ddot{\alpha}(\chi, t(\chi)) \delta t \\
\frac{d \delta t}{d \chi} & = -\dot{\alpha}(\chi, t(\chi))  \delta t
\end{align}
which, once solved along the geodesic $t_1=t(\chi, t_0)$,  provide a numerical estimate of the redshift drift.

\bibliography{reddrift}

\end{document}